# Phase Coherent Link of an Atomic Clock to a Self-Referenced Microresonator Frequency Comb

Pascal Del'Haye[1,2,*], Aurélien Coillet[1,†], Tara Fortier[1], Katja Beha[1], Daniel C. Cole[1], Ki Youl Yang[3], Hansuek Lee[3,‡], Kerry J. Vahala[3], Scott B. Papp[1], Scott A. Diddams[1,§]

[1]*National Institute of Standards and Technology (NIST), Boulder, CO 80305, USA*
[2]*National Physical Laboratory (NPL), Teddington, TW11 0LW, United Kingdom*
[3]*T. J. Watson Laboratory of Applied Physics, California Institute of Technology, Pasadena, CA 91125, USA*

**The counting and control of optical cycles of light has become common with modelocked laser frequency combs [1, 2]. But even with advances in laser technology [3], modelocked laser combs remain bulk-component devices that are hand-assembled. In contrast, a frequency comb based on the Kerr-nonlinearity in a dielectric microresonator [4, 5] will enable frequency comb functionality in a micro-fabricated and chip-integrated package suitable for use in a wide-range of environments. Such an advance will significantly impact fields ranging from spectroscopy and trace gas sensing [6-8], to astronomy [9], communications [10, 11], atomic time keeping [12, 13] and photonic data processing. Yet in spite of the remarkable progress shown over the past years [14-18], microresonator frequency combs ("microcombs") have still been without the key function of direct f-2f self-referencing [2] and phase-coherent frequency control that will be critical for enabling their full potential. Here we realize these missing elements using a low-noise 16.4 GHz silicon chip microcomb that is coherently broadened from its initial 1550 nm wavelength and subsequently f-2f self-referenced and phase-stabilized to an atomic clock. With this advance, we not only realize the highest repetition rate octave-span frequency comb ever achieved, but we highlight the low-noise microcomb properties that support highest atomic clock limited frequency stability.**

Focused research on microcombs has led to their realization in a growing range of materials and platforms [4, 5, 14, 19-26]. Together with the device development, there has been work aimed at the frequency control of microcombs [13, 17, 27-29], which is critically needed for many applications. To date, however, full frequency stabilization of a self-referenced microcomb has not been demonstrated. Detection of microcomb offset frequencies has been problematic and could only be achieved using an additional reference frequency comb [27, 30] or 2f-3f detection with transfer lasers [17]. Self-referencing is of particular importance because it provides direct measurement and control of the offset frequency of the microcomb, and is the key to forming a straightforward microwave-to-optical link. This is most simply implemented with an octave-spanning spectrum and a nonlinear f-2f interferometer to compare long and short wavelengths of the comb [31]. Our work highlights coherent spectral broadening to an octave bandwidth to demonstrate f-2f self-referenced stabilization of the microcomb offset frequency and its mode-spacing at levels provided by state-of-the-art atomic clocks.

Figure 1a shows the experimental setup for self-referencing of a microcomb. A tunable external cavity diode laser is amplified to ~100 mW and coupled into a fused silica microdisk resonator via a tapered optical fibre [32]. The microdisk resonator [33] has a diameter of ~4 mm and a corresponding free spectral range of 16.4 GHz. The coupled linewidth of the resonator mode family for comb generation is 1.7 MHz, which yields a quality factor of $Q = 1.1 \times 10^8$ at the pump laser wavelength of 1550 nm. The microresonator generates a phase locked microcomb [34, 35] with part of the comb light being detected by a fast photodiode to measure the repetition rate. The rest of the comb light is optimized in phase and amplitude using a liquid crystal based spatial light modulator [22, 36] in order to generate the shortest possible pulse at the input of a highly nonlinear fibre (HNLF 1 in Fig 1a, input power ~150 mW). Note that this phase and amplitude optimization could be removed when using a single soliton microcomb generator as demonstrated in $Si_3N_4$ and $MgF_2$ [16, 41]. After another step of quadratic dispersion compensation and amplification to a ~4 W average optical power, the 16.4-GHz pulse train of <200 fs pulses is sent into a second hybrid nonlinear optical fibre [37] that broadens the optical spectrum to an octave (Fig 1b). The octave spanning spectrum is sent to an f-2f interferometer for generation of the carrier envelope offset beat note [31, 38]. The f-2f interferometer includes a variable time delay for the long wavelength part of the spectrum in order to achieve a temporal overlap with the short wavelength end of the spectrum. Nonlinear frequency doubling of the 2.22 μm spectral region to 1.11 μm is achieved in a 10 mm long periodically poled lithium niobate crystal. Both the measured repetition rate $f_{rep}$ and carrier envelope offset frequency $f_{ceo}$ are referenced to a hydrogen maser frequency reference. Feedback to power and frequency of the pump laser allows stabilization of both $f_{ceo}$ and $f_{rep}$ of the microcomb. Figure 1c shows a photograph of the 4-mm-diameter fused silica disk with tapered optical fibre that has been used in this experiment.

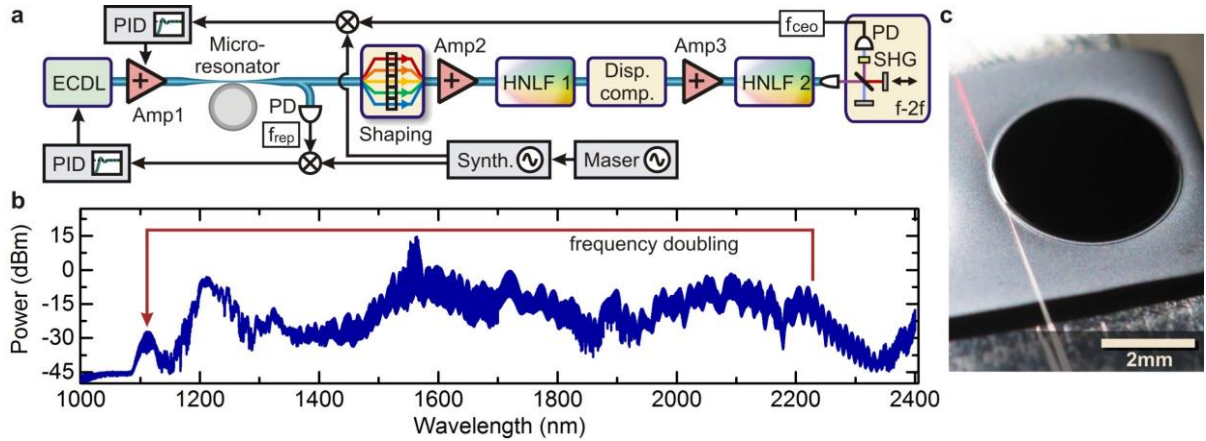

**Figure 1 | Experimental setup for f-2f self-referencing of a microcomb. a**. The microcomb is generated by an amplified external cavity diode laser (ECDL) and phase optimized for the generation of Fourier limited pulses shorter than 200 fs. Subsequent amplification and broadening in highly nonlinear fibre (HNLF) generates an octave spanning comb spectrum and enables the measurement of the carrier envelope offset frequency ($f_{ceo}$) using an f-2f interferometer. Repetition rate $f_{rep}$ and carrier envelope offset $f_{ceo}$ of the microcomb can be stabilized to an atomic clock (hydrogen maser). Amp: erbium doped fibre amplifier, PD: photodiode, PID: proportional-integral-derivative controller, "Shaping": liquid-crystal-based spatial light modulator, SHG: second harmonic generation. **b**. Octave spanning microcomb spectrum after nonlinear broadening. **c**. Photograph of the 4-mm-diameter $SiO_2$ disk used in the experiment.

Figure 2a shows the microcomb spectrum around 1.11 μm wavelength together with the frequency doubled light from 2.22 μm measured with an optical spectrum analyser at a resolution bandwidth of 5 GHz. This resolution bandwidth allows to resolve the modes of the microcomb and gives an estimate of the carrier envelope offset frequency of 7.5 GHz (frequency spacing between modes at 1.11 μm and frequency doubled modes). The corresponding carrier envelope offset beat note ($f_{ceo}$) detected by interfering the fundamental and doubled light on a fast photodiode is shown in Fig 2b, exhibiting 30 dB signal to noise at a resolution bandwidth of 100 kHz. Figure 2c shows a 25 minute recording of the carrier envelope offset frequency and the corresponding Allan deviation. The microcomb $f_{ceo}$ is stable at the 1% level over many weeks (when using the same pump mode and similar comb states). Significantly, we note that the free-running fractional stability of the microcomb $f_{ceo}$ is comparable to the best solid state laser frequency combs, even with no special attention given to environmental isolation at this point.

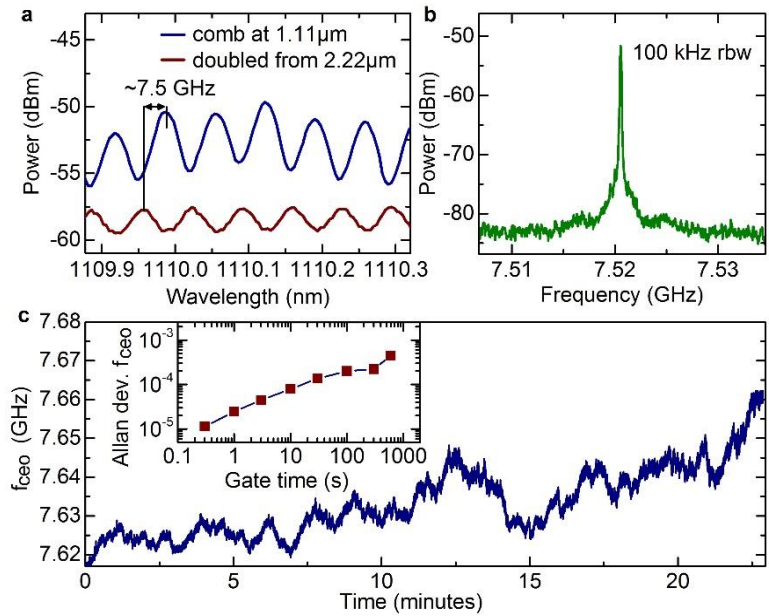

**Figure 2 | Carrier envelope offset frequency measurement. a**. Measurement of microcomb modes around 1.11 μm wavelength together with frequency doubled comb modes originating from 2.22 μm wavelength. The carrier envelope offset frequency of

~7.5 GHz can be resolved as offset between the fundamental and frequency doubled comb modes (spectrum analyser resolution bandwidth ~5 GHz). **b**. Electronic spectrum of the carrier envelope offset beat note with signal to noise >30 dB. **c**. Time series measurement of the free running carrier envelope offset frequency. The drift is attributed to lab temperature fluctuations and pump laser power/frequency drifts. The inset shows the Allan deviation of $f_{ceo}$ normalized to the 16.4 GHz repetition rate.

In order to verify the measurement of the actual carrier envelope offset frequency of the microcomb, we compare the frequency of the pump laser with an independent measurement using a self-referenced Er:fiber frequency comb. The optical frequency of the pump laser is determined from the microcomb's repetition rate $f_{rep}$ and carrier envelope offset frequency $f_{ceo}$

$$f_{pump1} = f_{ceo} + n \times f_{rep} \quad , \quad (1)$$

with $n$ being an integer number corresponding to the mode number of the pump laser. In addition, the pump laser frequency measured with the Er:fiber frequency comb with carrier envelope offset frequency $f_{ceoFC}$ and repetition rate $f_{repFC}$ yields:

$$f_{pump2} = f_{ceoFC} + m \times f_{repFC} + f_{beat} \quad . \quad (2)$$

Here, $m$ is the mode number of the closest fibre comb mode to the pump laser and $f_{beat}$ is the measured beat note frequency between the pump laser and the closest fibre comb mode. Figure 3 shows an illustration of the measurement and a plot of the measured differences between $f_{pump1}$ and $f_{pump2}$. Note that this measurement was acquired with microcomb repetition rate and carrier envelope offset frequency free running. The microcomb's carrier envelope offset frequency has been measured using the peak finder function of a Maser-referenced electronic spectrum analyser. This allowed us to measure a carrier envelope offset frequency beat note with very small signal to noise (< 5 dB) with a measurement rate exceeding 100 samples per second at a resolution bandwidth of 20 kHz in FFT mode (long term direct $f_{ceo}$ counting measurements were too unreliable because of insufficient signal to noise). Note that the scatter of ~160 kHz of the frequency measurement is due to the timing synchronization between measuring the free running $f_{ceo}$ and $f_{rep}$ of the microcomb. The agreement between the measurements of the pump laser frequency with the microcomb and the fibre laser comb is 670 Hz +/- 1.4 kHz, limited by the measurement time. Stabilization of the microcomb significantly improves the measurement accuracy, which is shown in the remaining part of the paper.

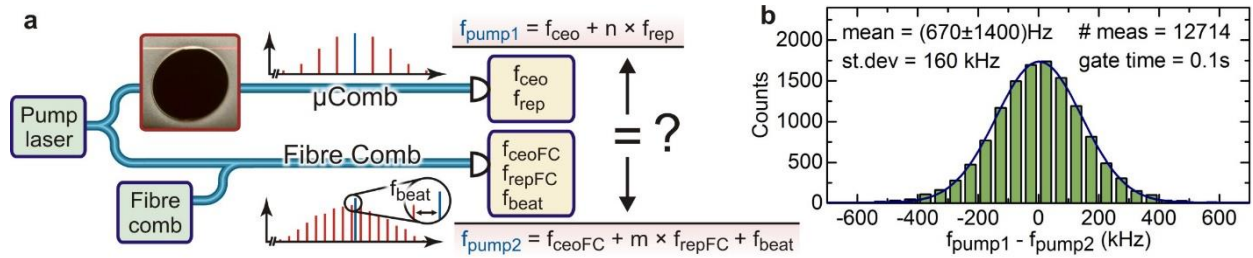

**Figure 3 | Absolute optical frequency measurement and out-of-loop validation. a**. Setup diagram. The free running pump laser frequency is determined by measuring the microcomb carrier envelope offset frequency and repetition rate. Simultaneously it is measured with a conventional fibre laser frequency comb. **b**. Histogram showing the agreement of the free running pump laser frequency measurements at a measurement time limited level of 670 Hz +/- 1.4 kHz.

Figure 4 shows a block diagram of a setup that is used to stabilize the carrier envelope offset frequency of the microcomb. The initial carrier envelope offset beat note at 7.5 GHz is amplified and mixed down to ~640 MHz using a hydrogen maser referenced synthesizer. In a next step, the signal is divided by 64 and phase locked to a maser-referenced 10 MHz signal. The error signal for the phase locked loop is generated in a digital phase comparator and fed into a PID controller. Control of the carrier envelope offset frequency is achieved by actuating on the power (via Amp1, Fig 1a) or frequency of the pump laser. Both these actuators change the intracavity power and detuning between the microresonator mode and pump laser. This effectively changes both comb spacing and carrier envelope offset frequency via fast thermal effects and Kerr effect [27]. Figure 4b,c show in-loop measurements of the stabilized carrier envelope offset frequency beat notes (after mixing down to 640 MHz and dividing by 64) for stabilization via pump power and pump frequency respectively. The sharp carrier peaks indicate phase-locked control relative to the 10 MHz maser signal.

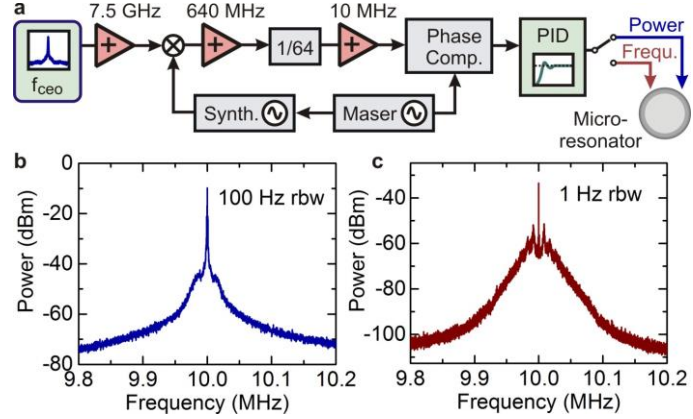

**Figure 4 | Carrier envelope offset frequency stabilization. a**. Block diagram of the electronic setup for microcomb carrier envelope offset stabilization. Panel **b,c** show the stabilized carrier envelope offset beat note using pump power (**b**) and pump frequency (**c**) as actuator.

A full stabilization of the microcomb is achieved by using the pump laser frequency to control the comb spacing and the pump power to control the carrier envelope offset frequency (this configuration has been chosen in order to minimize cross-talk between the actuators). Figure 5a shows a simultaneous frequency counter measurement of both $f_{ceo}/64$ and $f_{rep}$ compared to their setpoints. Note that the employed phase locked microcomb state only allowed for a limited variation in pump power and pump frequency, which leads to a limited "capture-range" of the phase locked loop (this could be improved by better environmental control and by using more robust comb states). Nevertheless, we were able to stabilize both degrees of freedom of the microcomb simultaneously to a sub-Hertz residual noise level (at 1 second gate time). Based on these measurements we calculate the absolute pump laser stability with a standard deviation of 620 Hz at a 100 ms gate time, which is consistent with the stability of the employed hydrogen maser references. Figure 5c shows the Allan deviation for $f_{ceo}$, $f_{rep}$ and the pump laser frequency calculated as $f_{pump} = f_{ceo} + 11709 \times f_{rep}$. It shows that the pump laser stability is mostly limited by repetition rate fluctuations due to the large multiplicative factor. At 1 second gate time the carrier envelope offset frequency is stabilized to a sub-Hz-level and only contributes at the $10^{-14}$ level to the optical carrier stability. Figure 5d,e show the stabilized and maser-referenced beat notes of $f_{ceo}/64$ and $f_{rep}$, respectively. Notice that the carrier envelope offset frequency exhibits additional side lobes at a frequency offset of ~80 kHz, which arise from cross-talk with the repetition rate phase locked loop.

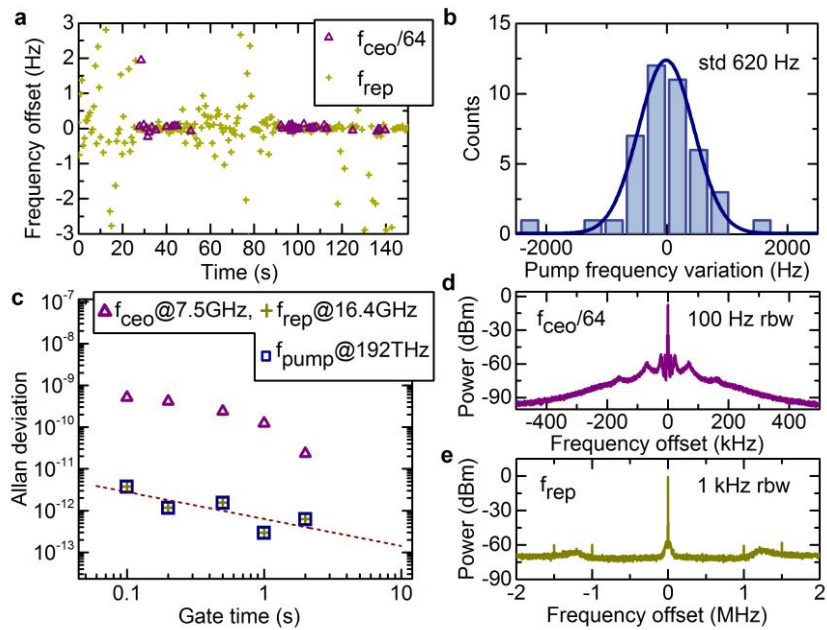

**Figure 5 | Stabilizing a self-referenced microcomb to a hydrogen maser frequency reference. a**. Simultaneous stabilization of carrier envelope offset frequency and repetition rate. Plotted are the counted in-loop signals of $f_{ceo}/64$ and $f_{rep}$. In order to maintain stable comb generation the actuation range is reduced leading to a limited "catch range" of the frequency lock. **b**. Histogram of the measured pump laser frequency with a standard deviation of 620 Hz at 100 ms gate time. **c**. Allan deviations of $f_{ceo}$, $f_{rep}$, and $f_{pump}$. The pump laser frequency measurement is limited by the repetition rate stabilization, which is multiplied by the pump laser mode number of ~12000. The red dashed line is a least squares fit of the pump laser Allan deviation. **d**. Stabilized carrier envelope offset frequency beat note after mixing down and dividing by 64. **e**. Stabilized repetition rate beat note.

In conclusion, we have demonstrated f-2f self-referencing of a microresonator-based optical frequency comb at a repetition rate of 16.4 GHz. The carrier envelope offset frequency is controlled via the pump laser power and frequency. Moreover, we simultaneously stabilize the carrier envelope offset frequency and repetition rate to a hydrogen maser based atomic clock frequency reference. A self-referenced measurement of the pump laser frequency with an out-of-loop comparison with a conventional frequency comb confirms the viability of microcombs for metrology applications. Looking forward, our demonstration of external broadening of the microcomb spectrum can take advantage of chip-integrated highly nonlinear waveguides [39, 40] to realize microphotonic self-referenced optical frequency comb systems. Taken together with progress in the generation of low-noise soliton states [16, 41], advanced dispersion engineering [42-44, 24, 18], and octave-span dispersive wave generation [45], these results highlight the future direction of chip-integrated microcombs as phase coherent microwave-to-optical links.

**Acknowledgements**

This work is supported by NIST, NPL, Caltech, the DARPA QuASAR program, the AFOSR and NASA. PD thanks the Humboldt Foundation for support. DCC acknowledges support from the NSF GRFP under Grant No. DGE 1144083.

**Author contributions**

PD, SBP and SAD conceived the experiments. PD and AC designed and performed the experiments. TF contributed to the $f_{ceo}$ stabilization. KB and DCC contributed to the nonlinear spectral broadening. KYY, HL, and KJV provided the microresonator. PD and SAD prepared the manuscript with input from all co-authors.

**Competing financial interests:** The authors declare no competing financial interests.

*pascal.delhaye@npl.co.uk

†current affiliation: Laboratoire Interdisciplinaire Carnot de Bourgogne, UMR CNRS 6303, 9 Avenue Alain Savary, 21078 Dijon, France

‡current affiliation: Graduate School of Nanoscience and Technology, Korea Advanced Institute of Science and Technology, Daejeon 305-701, South Korea

§scott.diddams@nist.gov